\documentclass[prb,twocolumn,showpacs,floatfix]{revtex4}
\usepackage{graphicx}
\usepackage{dcolumn}
\usepackage{bm}
\begin{document}
\title{Direct evidence of nuclear spin waves in Nd$_2$CuO$_4$ by high-resolution neutron-spin-echo spectroscopy}
\author{Tapan Chatterji$^1$, Olaf Holderer$^2$ and Harald Schneider$^2$}
\address{$^1$Institut Laue-Langevin, B.P. 156, 38042 Grenoble Cedex 9, France\\
$^2$JCNS Outstation FRM II, Forschungszentrum J\"ulich, J\"ulich, Germany$^2$
}
\date{\today}
\begin{abstract}
We investigated the dispersion of nuclear spin waves in Nd$_2$CuO$_4$ by using neutron spin-echo spectroscopy at millikelvin temperatures. Our results show unambiguously the existence of dispersion of nuclear spin waves in Nd$_2$CuO$_4$ at T = 30 mK. A fit of the dispersion data with the spin wave dispersion formula gave the Suhl-Nakamura interaction range to be of the order of 10 {\AA}.
\end{abstract}
\maketitle
The coupling of nuclear spins through Shul-Nakamura indirect interaction gives rise to nuclear spin excitations \cite{keffer66}. Each nuclear spin sees the electronic spin on its own ion through the effective hyperfine coupling $A{\bf I.S}$, where A is the hyperfine parameter, {\bf I} is the nuclear spin and $\bf {S}$ is the electronic spin. The electronic spins of all the ions are coupled by exchange interaction. An interaction of the nuclei therefore arises via the low-lying excited states (spin waves) of the electronic systems as intermediate states. That is to say a nuclear spin excites a spin wave through the hyperfine coupling, another nuclear spin causes it to be reabsorbed through its hyperfine coupling. This process gives rise to the so-called Shul-Nakamura indirect interaction \cite{suhl58,nakamura58}. Although the polarised nuclear spin system is far from being in perfectly ordered state, it does posses long range order of the average oriented nuclear spin because of the long range of the Shul-Nakamura interaction; and therefore as pointed out and justified by de Gennes et al. \cite{degennes63}, there exists a spin-wave like spectrum of excitations of the nuclear spin system. Word et al. \cite{word77} calculated the relevant neutron scattering cross-sections from electronic and nuclear spin systems coupled by Suhl-Nakamura interaction. There exists indirect evidence of the nuclear spin-waves, which cause lowering of the antiferromagnetic resonance (AFR) and nuclear magnetic resonance (NMR) frequencies at low temperatures \cite{witt64}. Nuclear spin waves are expected to have energies about three orders of magnitude lower than those of electronic spin waves and should be in the $\mu$eV range. The nuclear spin wave spectrum should have dispersion at lower q range inversely proportional to the range b of the Shul-Nakamura indirect interaction. Kurkin and Turov \cite{kurkin88} have considered the properties of nuclear spin waves and have compared them with those of electronic spin waves. Owing to the large value of Suhl-Nakamura interaction range there can be still a strong correlations in the motion of nuclear spins, even when they are in a disordered paramagnetic state. In the electronic case such strong correlations are seen only in low-dimensional spin systems. We already commented that the energy scale of nuclear spin waves is three orders of magnitude less that of the electronic spin waves or magnons. There are differences in the dispersion laws for nuclear spin waves and magnons due to the difference in the coordinate dependence of the Suhl-Nakamura and exchange interactions. Whereas the Fourier spectrum of the short-range exchange interaction contains all the wave vectors $\bf {k}$ from the Brillouin zone, in the Fourier spectrum of the long-range Suhl-Nakamura interaction the components with k much greater than $r_0^{-1}$  ($r_0$ is the Suhl-Nakamura interaction range) are sparsely represented. There is another very important difference between the properties of nuclear and electronic spin waves. The spin wave dispersions in ferro- and antiferromagnets are very much different in electronic spin systems, whereas in the case of nuclear spin waves they are very similar. All effects  associated with nuclear spin waves are very similar but much stronger in antiferromagnets due to the drastic difference in the Shul-Nakamura interaction range. 

    There have been already some NMR investigations \cite{witt64,lee63,minkie66,welsh67,shaltiel64}, which probe the nuclear spin waves indirectly and are essentially limited to q = 0. These studies have been done on Mn-based compounds., MnCO$_3$, CsMnF$_3$, CsMnCl$_3$ and RbMnF$_3$. The $^{55}$Mn nucleus (spin $I = 5/2$) has a large hyperfine constant $A = 600$ MHz and 100\% abundance of the magnetic isotope. Nuclear spin waves have therefore been studied on Mn-based compounds by the NMR technique. However, neutron scattering is the only microscopic probe that is in principle capable of measuring nuclear magnetic excitations in the $(Q,\omega)$ space. But in order to probe nuclear spin waves it is to have very good energy as well as $q$ resolution at the same time. The neutron spin-echo spectrometers are the only instruments that satisfy such stringent requirements. Another important factor is of course suitably large spin dependent scattering cross sections. Nd-compounds satisfy such requirements. Out of seven naturally occurring isotopes $^{143}$Nd and $^{145}$Nd have spin $I =7/2$ with natural abundances 12.18\% and 8.29\%, respectively and the spin dependent scattering cross sections are large. We have therefore investigated nuclear spin excitations in the well-known parent compound Nd$_2$CuO$_4$ of the electron-doped superconducting compounds Nd$_{2-x}$Ce$_x$CuO$_4$. A large well-characterized single crystal of Nd$_2$CuO$_4$ was available to us. Hyperfine induced nuclear spin ordering in Nd$_2$CuO$_4$ below about 200 mK has been reported by Chattopadhyay and Siemensmeyer \cite{chatt95} from their neutron diffraction measurements. Chatterji and Frick \cite{chatt00} have reported observation of nuclear spin excitations in Nd$_2$CuO$_4$ from their inelastic neutron scattering investigation with a back-scattering spectrometer. However the dispersion of the nuclear spin waves that is expected to occur at a very small $q$ could not be measured because the back-scattering spectrometer did not have the required $q$ resolution. During the present investigation we attempted to determine the dispersion of nuclear spin waves in Nd$_2$CuO$_4$ on a neutron spin echo spectrometer and have succeeded in showing the existence of the dispersion of nuclear spin waves at $T = 40$ mK. The results of this investigation are reported in the following.
\begin{figure}
\resizebox{0.5\textwidth}{!}{\includegraphics{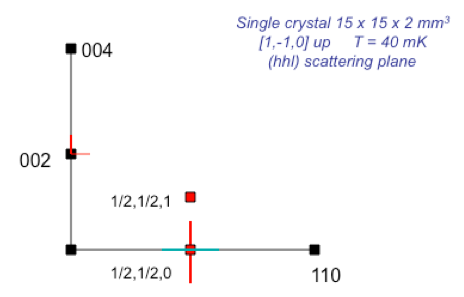}}
\caption {(Color online) $(hhl)$ reciprocal scattering plane and the scan directions have been  illustrated schematically.  }
\label{reciprocal}
\end{figure}

\begin{figure}
\resizebox{0.5\textwidth}{!}{\includegraphics{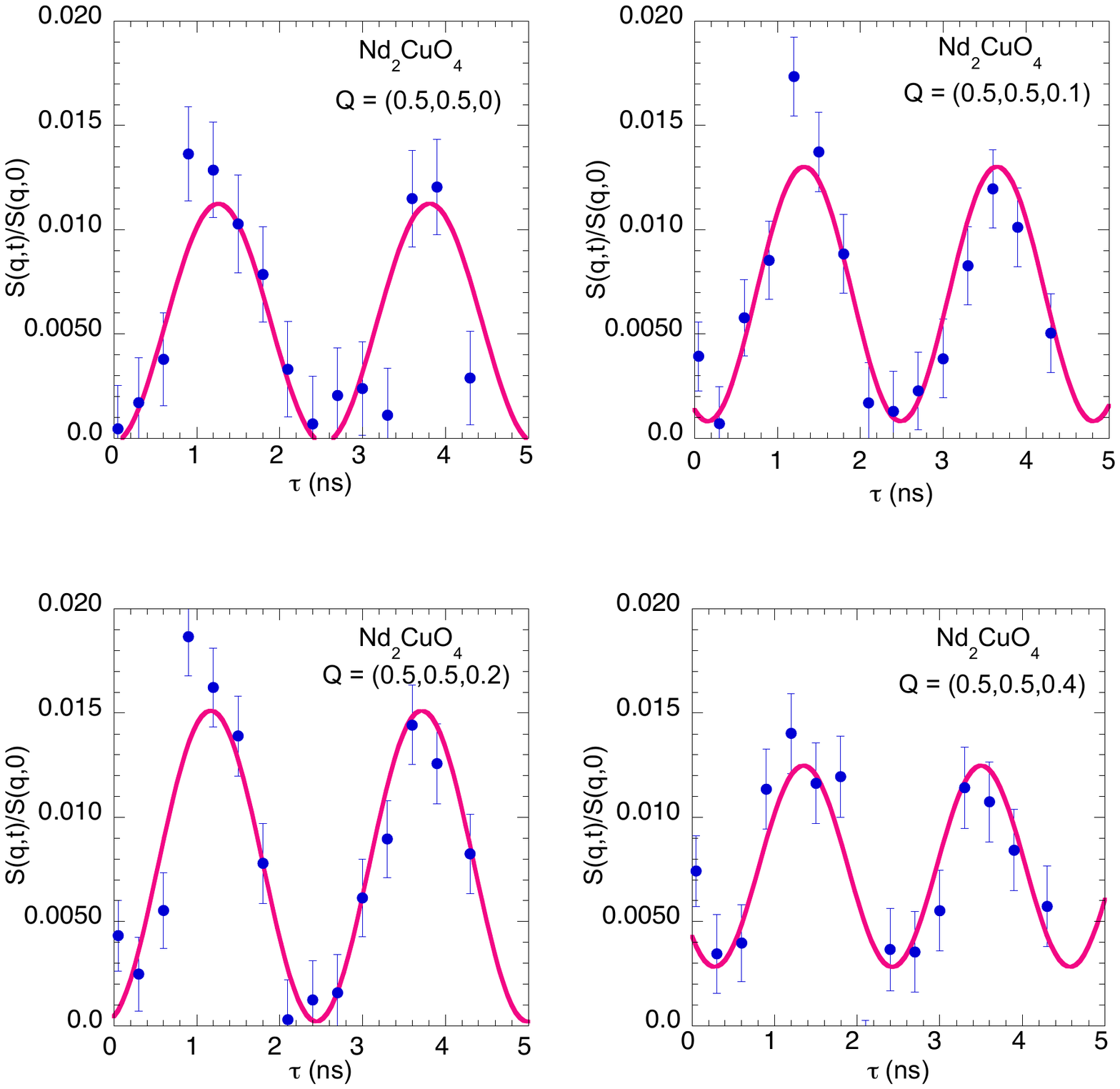}}
\caption {(Color online) Ð $S(q,\tau)/S(q,0)$ as a function of the Fourier time $\tau$  at $Q = ( \frac{1}{2}, \frac{1}{2}, 0)$ , $Q = ( \frac{1}{2}, \frac{1}{2}, 0.1)$, $Q = ( \frac{1}{2}, \frac{1}{2}, 0.2)$, and at $Q = ( \frac{1}{2}, \frac{1}{2}, 0.4)$  with the least-squares fit results with a sine function.  }
\label{scans}
\end{figure}

\begin{figure}
\resizebox{0.5\textwidth}{!}{\includegraphics{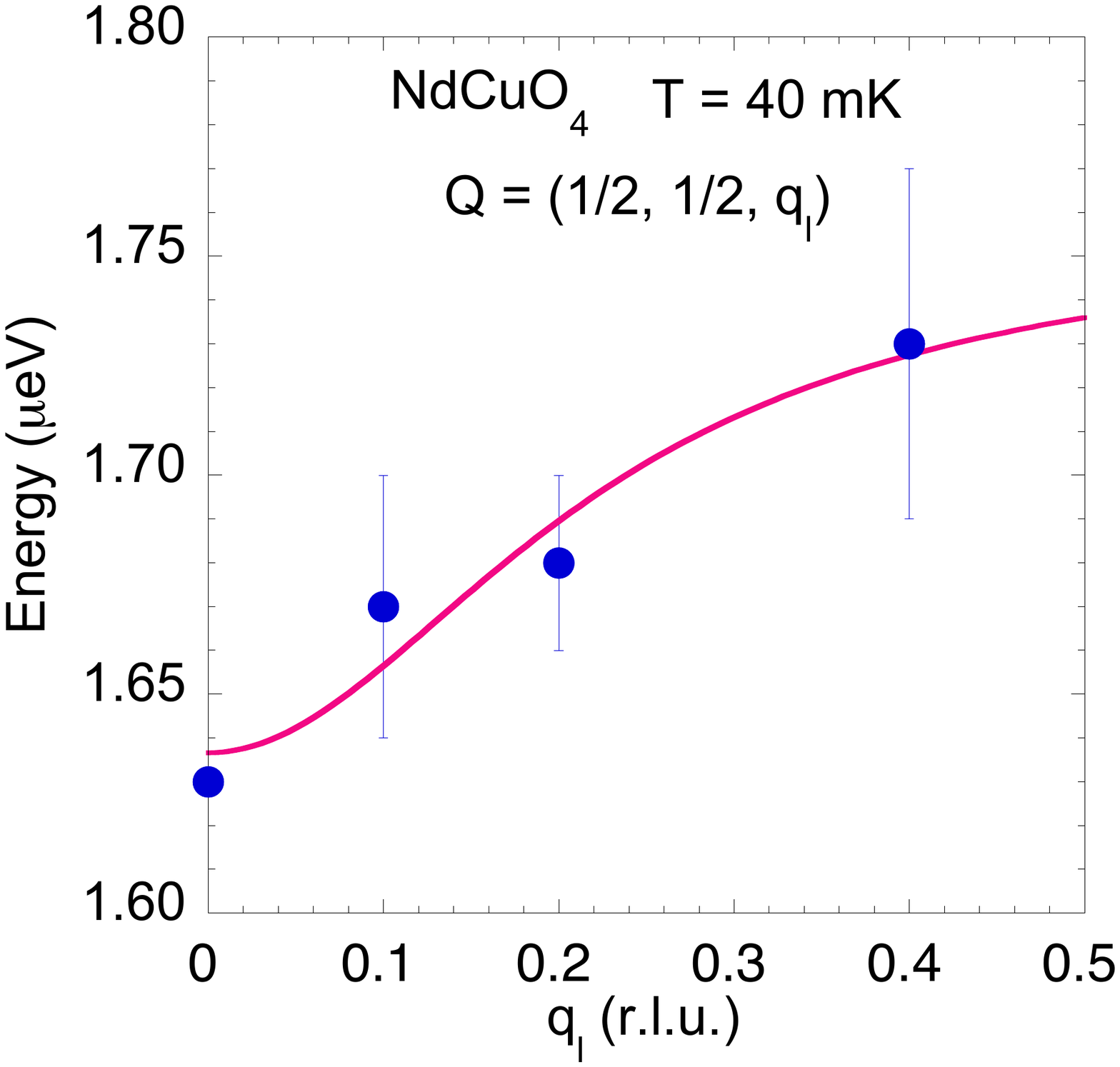}}
\caption {(Color online) Dispersion of the nuclear spin waves in Nd$_2$CuO$_4$ along [001]. The red curve shows the least-squares fit  of the dispersion data with the nuclear spin wave model explained in the text. }
\label{dispersion}
\end{figure}

    The experiment was carried out at the J-NSE spetrometer \cite{holderer08,monken97} at the FRM II . The sample was mounted in a Kelvinox cryostat with $^3$He-$^4$He dilution insert and was cooled down to 40 mK. The sample temperature was kept at 40 mK throughout the experiment. The J-NSE was equipped with a rotating sample stage and the second arm could be turned in the horizontal plane with the origin of rotation at the sample position. However, tilting the sample was not possible due to the absence of the tilting goniometers in the spectrometer. Therefore the crystal was oriented before with its $[1,-1,0]$ crystallographic axis vertical such that the scattering plane was $(hhl)$ enabling one to access the nuclear magnetic Bragg peak at $( \frac{1}{2}, \frac{1}{2}, 0)$  and measure the dispersion along $[110]$ and $[001]$. Fig. \ref{reciprocal} shows schematically the reciprocal scattering $hhl$ plane and scan directions. The beam was collimated by the distance between the neutron guide exit ($60\times 60$ mm$^2$) and the sample ($10 \times 10$ mm$^2$) and has a divergence of about 0.5$^\circ$. A 10\% velocity selector provided a neutron beam with a wavelength of $5 \pm 0.25$ {\AA}. A two-dimensional $^3$He-multidetector at a distance of 4.33 m from the sample accepts an angular range of $\pm 2^\circ$ from the central point.  The normalized intermediate scattering function $S(q,\tau)/S(q,0)$ has been measured in the standard configuration (including  $\pi$-flipper) \cite{monken97}. 
    
    A reference measurement with an elastic scatterer (a stochastic TiZr alloy) has been
done for $|q|=| \frac{1}{2}  \frac{1}{2} 0 |$ and the same set of Fourier times. It measures the
instrumental resolution. Compared to measurements of S(q,$\omega$), where resolution
corrections involve a Fourier transform ( unfolding of the spectra with the
instrumental resolution function), this correction is a simple division in neutron
spin echo spectroscopy due to the intrinsic Fourier transform $S(q,\tau)/S(q,0)$ of the NSE technique. 
The q-resolution in the direction of $\vec{q}$ is determined by the divergence of
the beam and the wavelength spread $\Delta \lambda / \lambda =0.1$, while
perpendicular to it only the beam divergence is the mayor contribution, this
elongated instrumental resolution favours measurements in the $[1, 1, X]$-direction. The oscillations measured in this experiment always started at a minimum, which indicates that the resulting scattered intensity is a sum of an elastic non-spin-flip and the oscillating spin-flip processes. The ratio of the two determines the y-axis offset of the oscillation. 

The spin flip scattering flips all spin directions (like the $\pi$-flipper). Similar
contributions of residual elastic coherent scattering and the magnetic scattering
results in a completely depolarized elastic signal (no difference in scattered
intensity, whether  the instrumental $\pi$-flipper is on or off).  The inelastic
measurements show a minimum at t=0. When the magnetic signal goes to the first
minimum, only the residual elastic coherent contribution remains and gives rise to a
maximum of the oscillation.

   Only a limited number of scans could be measured due to measuring times of 2-3 days per scan. The evaluated detector regions avoided the borders of the detector and the Bragg-peak, if still visible on the detector. The oscillating signal could not be normalized as usual to the up- and down-intensity (all flippers off, only $\pi$-flipper on respectively), since the mixture of spin-flip and non-spin-flip scattering erases the differences between up- and down-configuration, but not the Fourier time dependence of the signal. The non-spin-flip contribution is an elastic (i.e. constant) background arising either from residual elastic scattering of the sample or/and the cryostat Al-windows. The intermediate scattering function has been fitted with the function $S(q,\tau)/S(q,0)=A\sin(\omega \tau)+B$, since NSE measures the cosine Fourier transform of the scattering function $S(q,\omega)$ the oscillating frequency is directly related to the energy of the excitation $E=\hbar \omega$. Here $A$ is the amplitude and $B$ is the background. The nominal position of the nuclear magnetic Bragg peak, $Q=(\frac{1}{2}, \frac{1}{2}, 0)$ has been set to the maximum intensity of the peak after the $\theta-2\theta$-scan. Measuring $S(q,\tau)/S(q,0)$ at this position and evaluating the detector in the vicinity of the peak lead to a very noisy and weak oscillation, hence large errors are encountered for this setup. Moving away from the peak by rotating the sample, an oscillation clearly could be identified with a frequency of $f=2.62$ ns$^{-1}$ $(\hbar \omega=1.73 \mu$eV) where $\omega=2\pi f$. This oscillation frequency was rather independent on the details of the evaluation and the exact size of the patch on the detector evaluated.  By turning the sample, one takes advantage of the higher resolution in the direction perpendicular to $Q$, but still averages in the other direction over some $|q|$ due to the 10\% wavelength spread, which limits the possibility to go to small $|q|$ in the dispersion relation. 

Fig. \ref{scans} shows $S(q,\tau)/S(q,0)$ as a function of the Fourier time $\tau$ at $Q = ( \frac{1}{2}, \frac{1}{2}, 0)$ , $Q = ( \frac{1}{2}, \frac{1}{2}, 0.1)$, $Q = ( \frac{1}{2}, \frac{1}{2}, 0.2)$, and at $Q = ( \frac{1}{2}, \frac{1}{2}, 0.4)$  with the least-squares fit results with a sine function and Fig. \ref{dispersion}  shows the resulting dispersion. We fitted the dispersion with the expected dispersion for a nuclear spin system [7] coupled by Suhl-Nakamura interaction 
\begin{equation}
  E(q)=E_{s}-E_p\left[1+(qr_0)^2\right]^{-1}                                       
\end{equation}
where  $E(q)$ is the energy of nuclear spin excitation at the momentum transfer vector $q$, $E_{s}$ is the saturation value of energy, $E_p$ is the amount of dispersion corresponding to the frequency pulling $\omega_p$ in NMR  experiment and $r_0$  is the Suhl-Nakamura interaction range. The least-squares fit give $E_{s} = 1.76 \pm 0.5 \mu eV$, $E_p = 0.12 \pm 0.05 \mu eV$ and $r_0 = 5 \pm 3$ direct lattice units. Due to the lack of enough data points and also due to the uncertainty of the experimental q values these fitted parameters have relatively large standard deviations. 

We note that the present results give the first inelastic-neutron-scattering evidence for the nuclear spin waves. However the results are certainly very preliminary and we still lack of enough data points for determining the dispersion and its temperature dependence. Despite these shortcoming of the present results we note that our results gives the corresponding frequency pulling in NMR experiment at q = 0 to be 0.12 $\mu$eV.  Unfortunately no such NMR experiment has been done so far. The Suhl-Nakamura interaction range turns out to be $5c/2\pi \approx 10$ {\AA}. This value is however much less than expected from Suhl-Nakamura interaction range (100 {\AA} or so) speculated theoretically \cite{degennes63,kurkin88}. However the fitted data for the Suhl-Nakamura interaction range has a very large standard deviation.  It is to be noted that the non-availability of the tilting goniometers in the spectrometer has caused uncertainty in the determination of the q values. So although we have unambigously demonstrated the existence of the dispersion of nuclear spin waves in Nd$_2$CuO$_4$ the determination of the Suhl-Nakamura interaction range is not yet accurate enough. A higher q-resolution would be desirable for investigating in more detail the
dispersion relation close to the nuclear magnetic Bragg peak and for the precise determination of the Nakamura interaction range . This could be achieved with a time of flight NSE spectrometer (e.g. the SNS-NSE \cite{ohl12}), where one is not
limited to the 10\% velocity selector for q-resolution but can adapt a posteriori the
resolution by choosing the desired number of time frames for evaluation.

In conclusion we have shown unambiguously from our inelastic neutron scattering investigation with a neutron spin-echo spectrometer the existence of dispersion of nuclear spin waves in Nd$_2$CuO$_4$ at $T = 30$ mK.

    We thank M. Monkenbusch and L.P. Regnault for critical discussions. This research project has been supported by the European Commission under the 7th
Framework Programme through the "Research Infrastructures" action of the
"Capacities" Programme, NMI3-II Grant number 283883.

\end{document}